\documentclass[aps,pra,amsfonts,amssymb,amsmath,showpacs,floatfix,nofootinbib,groupedaddress,superscriptaddress,citesort,twocolumn]{revtex4}

\usepackage{amssymb,amsmath,amsthm}

\theoremstyle{definition}

\theoremstyle{remark}

\usepackage{subfigure}
\usepackage{mathrsfs}
\usepackage{longtable}
\usepackage{amsfonts}
\usepackage{amstext}
\usepackage[usenames]{xcolor}
\usepackage[dvips]{graphicx}
\def\qed{\leavevmode\unskip\penalty9999 \hbox{}\nobreak\hfill
     \quad\hbox{\leavevmode  \hbox to.77778em{%
              \hfil\vrule   \vbox to.675em%
               {\hrule width.6em\vfil\hrule}\vrule\hfil}}
     \par\vskip3pt}

\begin{document}
\title{The optimal approximations of available states and a triple uncertainty  relation}
\author{Xiao-Bin Liang}
\email{liangxiaobin2004@126.com}
\affiliation{School of Mathematics and Computer science, Shangrao Normal University,
 Shangrao 334001, China}
\author{Bo Li}
\email{libobeijing2008@163.com}
\affiliation{School of Mathematics and Computer science, Shangrao Normal University,
 Shangrao 334001, China}

\author{Liang Huang}
\affiliation{Department of Computer and Software Engineering, Wonkwang University, Iksan 54538, Korea
}

 \author{Biao-Liang Ye}
\affiliation{Quantum Information Research Center, Shangrao Normal University, Shangrao 334001, China}
\author{Shao-Ming Fei}
\email{feishm@cnu.edu.cn}
\affiliation{School of Mathematical Sciences, Capital Normal University, Beijing 100048, China}
 \affiliation{Max-Planck-Institute
for Mathematics in the Sciences, 04103 Leipzig, Germany}

\author{Shi-Xiang Huang}
\email{hsx8154562@126.com}
 \affiliation{School of Mathematics and Computer science, Shangrao Normal University,
 Shangrao 334001, China}

\begin{abstract}
We investigate the optimal convex approximation of quantum state with respect to a set of available states. By isometric transformation, we have presented the general mathematical model and its solutions, together with a triple uncertainty equality relation. Meanwhile, we show a concise inequality criterion for decomposing qubit mixed states. The new results include previous ones as special cases. Our model and method may be applied to solve similar problems in high-dimensional and multipartite scenarios.
\end{abstract}
\maketitle

\section{Introduction}

In quantum information processing and computation, convex structures play important roles in the ensemble
of quantum states, quantum measurements and quantum channels. A typical convex structure problem is the quantum state discrimination which distinguishes a quantum state from a set of given  quantum states $\{|\Psi{_i}\rangle\}^{n}_{i=1}$  with prior probabilities $p_i$ satisfying $\sum^{n}_{i}p_i=1$, see \cite{Helstrom,Barnett12,Donghoon, Masoud}. Recently, the problem of optimal approximations of an unavailable quantum state to a set of available states
has been considered in \cite{Sacchi4,Liang5,Sacchi41,Theurer6}. For a given state $\rho$, the problem is recast to look for the least distinguishable states from $\{|\Psi{_i}\rangle\}_{i=1}^{n}$ such that distance between $\rho$ and the convex set $\sum^{n}_{i}p_i|\Psi{_i}\rangle\langle\Psi{_i}|$ is minimized \cite{Sacchi41}, whose solution is beneficial to the selection of available quantum resources \cite{Zi,T,Shang1}. Similar to the choice of distance measure for quantum coherence and quantum entanglement, we adopt here the trace norm as the measure of distance \cite{Kavan,Hu,Gao,Rana7,Bromley9,Shao,Streltsov10}.

An important question is how to choose the bases $\{|\Psi{_i}\rangle\}_{i=1}^{n}$.
In quantum information processing, generally
one concerns the availability of logical gates in preparing quantum states. From the perspective of resource theory, the so-called available states usually mean that they can be readily prepared and manipulated. In optical experiments, an obliquely placed polarizer transforms an input photon state to the eigenstate of a real quantum logic gate. If the half-wave plate is obliquely placed at $\pi/8$ with the horizontal axis, it constitutes a Hadamard gate \cite{Barnett13,Clarke14}. Hence it is meaningful to consider eigenstates of real quantum logic gates as the available bases, either from the perspective of experiment availability or the feasibility in state preparing. The uncertainty relation given in \cite{Robertson} is often termed as preparation uncertainty, which deals with the spread of measurement outcomes rather than the Heisenberg's original idea about measurement inaccuracies \cite{Friedland,Giuseppe,Sulyok,Fei}. Recently, a preparation uncertainty relation for three pairwise canonical observables was derived in \cite{K}, and a triple uncertainty relation was shown by geometric methods \cite{Gisin17}. In this paper, from the view of optimization, we obtain the triple constant $\tau$  and the corresponding tight boundary point by using the analytical method, as well as a triple uncertainty equality relation.

In \cite{Liang5,Sacchi41,Theurer6}, the authors considered the optimal convex approximations from a target qubit state to the eigenstate set of any two Pauli matrices or all three Pauli matrices. In this paper, we consider more generally cases: optimal approximations to eigenstates of real quantum logic gates.

\section{Optimal approximations of mixed states to two quantum logic gates}

Any qubit mixed states $\rho$ can be parameterized as
\begin{eqnarray}
\rho =\left(
\begin{array}{cc} 1-a
& k
\sqrt{a(1-a)}e^{-i\phi }\\
k \sqrt{a(1-a)}e^{i\phi }
& a  \\ \end{array} \right ), \;\nonumber
\end{eqnarray}
where $a \in [0,1]$, $\phi \in [0,2\pi]$ and $k\in [0, 1]$.

\textbf{Definition 2.1.} \cite{Hu,Gao,Rana7} ~~\emph{The trace norm  of a matrix $A$ is given by $\|A\|_1=\mathrm{Tr }\sqrt{A^\dag A}=\sum_i\sqrt{r_i}$, where $r_i$ are the eigenvalues of $A^\dag A $}.

Consider a set $S=\{|\Psi_i\rangle,i=1,2,...,n\}$, where $|\Psi_i\rangle$ are pure states. Let
$|0\rangle=(1,0)^\dag$ and $|1\rangle=(0,1)^\dag$ be the computational bases, where $\dag$ stands for
conjugation and transposition. A pure qubit state can be written as
$|\Psi\rangle=z_1|0\rangle+z_2|1\rangle$, $\mid z_1 \mid^2+\mid z_1 \mid^2 =1$, $z_1,z_2\in \mathbb{C}$. 
The corresponding density matrix is given by $|\Psi\rangle\langle\Psi|=(z_1|0\rangle+z_2|1\rangle)(z_1|0\rangle+z_2|1\rangle)^\dag$.
For convenience, denote $\rho_i=|\Psi_i\rangle\langle\Psi_i|$, $i=1,2,\ldots,n$.

\textbf{Definition 2.2.}~~\emph{The optimal convex approximation of a given state $\rho$ to the set $S$ is defined by
$$
D_{S}(\rho)=\mathrm{min}\{\parallel\rho-\sum_i p_i\rho_i\parallel_1\},~ S=\{|\Psi_i\rangle,i=1,2,...,n\},
$$
where the minimization takes over all possible probability distributions of $p_i$, $0\leq p_i\leq 1$, $\sum_i p_i =1$. We denote $S(\rho^{opt})$ the set of optimal convex approximations $\rho^{opt}=\sum_i p_i\rho_i$ such that $D_{S}(\rho)=\parallel\rho-\rho^{opt}\parallel_1$, $S(\rho^{opt})=\{\rho^{opt}|D_{S}(\rho)=\parallel\rho-\rho^{opt}\parallel_1\}$.}

The above definition is actually a mathematical distinction between quantum states. The use of trace norm is due to its significance in quantum coherence and entanglement measures \cite{Kavan,Hu,Gao,Rana7,Bromley9,Shao,Streltsov10}.  Now we study the optimal approximation of a given qubit state to the set of eigenstates of arbitrarily two different real quantum logic gates. A real quantum logic gate is of the forms, either
\begin{eqnarray}
U_\alpha =\left(
\begin{array}{cc} \cos\alpha
& \sin\alpha\\
\sin\alpha
& -\cos\alpha \\ \end{array} \right )~~ \;\mathrm{or}~~
V_\gamma =\left(
\begin{array}{cc} \cos\gamma
& -\sin\gamma\\
\sin\gamma
& \cos\gamma \\ \end{array} \right ).\nonumber
\end{eqnarray}
If $\alpha$ is taken to be $0,\frac{\pi}{4}$ and $\frac{\pi}{2}$, then $U_\alpha$ reduces to the important
Z-gate, Hadamard-gate and X-gate in quantum information processing, respectively.
We denote $\langle U_\alpha\rangle$ the expected value $\mathrm{Tr}(\rho U_\alpha)$. Then
$\langle \sigma_z \rangle$ and $\langle \sigma_x \rangle$ are the mean expectation values of the Pauli matrices $\sigma_x$(X-gate) and $\sigma_z$(Z-gate), namely, $\langle U_0\rangle=\langle \sigma_z \rangle$ and $\langle U_{\frac{\pi}{2}}\rangle=\langle \sigma_x \rangle$.
Since $\langle \sigma_z \rangle =1-2a$, $\langle \sigma_x \rangle=2k\sqrt{a(1-a)}\cos\phi$, we have
$\langle U_\alpha\rangle=\cos\alpha\langle \sigma_z \rangle+\sin\alpha\langle \sigma_x \rangle$.

The eigenvectors of $U_\alpha$ are $|\Psi_1\rangle=\cos\frac{\alpha}{2}|0\rangle+\sin\frac{\alpha}{2}|1\rangle$ and $|\Psi_2\rangle=\sin\frac{\alpha}{2}|0\rangle-\cos\frac{\alpha}{2}|1\rangle$. Similarly, we denote $|\Psi_3\rangle=\cos\frac{\beta}{2}|0\rangle+\sin\frac{\beta}{2}|1\rangle$ and $|\Psi_4\rangle=\sin\frac{\beta}{2}|0\rangle-\cos\frac{\beta}{2}|1\rangle$ the eigenvectors of $U_\beta$.
In addition, let $|\Psi_5\rangle=\frac{\sqrt{2}}{2}(|0\rangle+\sqrt{-1}|1\rangle)$ and $|\Psi_6\rangle=\frac{\sqrt{2}}{2}(|0\rangle-\sqrt{-1}|1\rangle)$.
Obviously, $|\Psi_5\rangle$ and $|\Psi_6\rangle$ are the eigenvectors of $\sigma_y $ (Y-gate), which
are also the eigenvectors of $V_\gamma(\gamma\neq 0,\pi)$.

The reasons for choosing above $\{|\Psi{_i}\rangle\}_{i=1}^{6}$ vectors as bases are due to the availability and manipulability of the corresponding gates in quantum state preparation, and that the mathematical forms of the three components of angular momentum for spin $1/2$ correspond to the X-gate ($\sigma_x$), Y-gate ($\sigma_x$) and Z-gate ($\sigma_x$), respectively. We will show that the bases $\{|\Psi{_i}\rangle\}_{i=1}^{6}$ and the triple uncertainty relation are intrinsically linked. Given that the eigenvectors of three Pauli matrices and the Hadamard-gate are all included in the eigenvectors of the real logic gates, the convex combinatory of the bases $\{|\Psi{_i}\rangle\}_{i=1}^{6}$ is a very general setting.

In the following, we consider the optimal convex approximation of the initial qubit states $\rho_c$ with respect to arbitrarily two real quantum logic gates. The available state sets are
$S_1=\{|\Psi_1\rangle,|\Psi_2\rangle,|\Psi_3\rangle,
|\Psi_4\rangle\}$, $S_2=\{|\Psi_3\rangle,|\Psi_4\rangle,|\Psi_5\rangle,|\Psi_6\rangle\}$ and $S_3=\{|\Psi_1\rangle,|\Psi_2\rangle,|\Psi_5\rangle,|\Psi_6\rangle\}$. Because the case $S_2$ is completely equivalent to the case $S_3$, we just need to consider $S_1$ and $S_2$.
Denote $S^{'}=\{|0\rangle,|1\rangle$, $|2\rangle=\cos\theta|0\rangle+\sin\theta|1\rangle$, $|3\rangle=\sin\theta|0\rangle-\cos\theta|1\rangle \}$, where $|2\rangle$ and $|3\rangle$ are two-dimensional superposition states, $\theta=\frac{(\beta-\alpha)}{2}$, $\beta>\alpha$ and $\theta \in(0,\pi)$. For the convenience of later discussion, let $U$ be an isometric transformation that rotates all vectors by an angle of  $\frac{\alpha}{2}$  counterclockwisely, namely,
\begin{eqnarray}
U=\left(\begin{array}{cc} \cos\frac{\alpha}{2}
& \sin\frac{\alpha}{2}\\
-\sin\frac{\alpha}{2}
& \cos\frac{\alpha}{2}\\ \end{array} \right ).\nonumber
\end{eqnarray}
We denote $\rho=U\rho_c U^\dag $ as $U\rho_c U^\dag $ is again a density matrix with parameters $a ,\phi$ and $k$. By symmetry we can set $a \in [0,\frac{1}{2}]$ and $\phi \in [0,  \frac{\pi}{2}]$.

\textbf{Lemma 2.3.}~~\emph{The problem of the optimal convex approximation of {$\rho_c$} with respect to $S_1$ is equivalent to that of $\rho$ with respect to $S^{'}$.}

\emph{Proof.}
Obviously, one has $U|\Psi_1\rangle=|0\rangle$, $\mathrm{and}~U|\Psi_2\rangle=-|1\rangle$, and
then $U|\Psi_2\rangle\langle\Psi_2|U^\dagger=|1\rangle\langle1|$.
Moreover, $U|\Psi_3\rangle=(\cos\theta|0\rangle+\sin\theta|1\rangle)=|2\rangle$ and  
$U|\Psi_4\rangle=(\sin\theta|0\rangle-\cos\theta|1\rangle)=|3\rangle.$ Therefore,
\begin{eqnarray}
\parallel U(\rho_c-\sum^4_{i=1}p_i\rho_i )U^\dag\parallel_1&=& \nonumber \\ \parallel U\rho_c U^\dag-\sum^4_{i=1} p_i U|\Psi_i\rangle\langle\Psi_i|U^\dag\parallel_1&=&\parallel \rho-\sum^3_{i=0} p_i |i\rangle\langle i|\parallel_1 ,\nonumber
\end{eqnarray} 
which proves that the optimal convex approximations of $\rho_c$ with respect to the set $S_1$ is equivalent to that of $U\rho_c U^\dag =\rho$  with respect to the set  $S^{'}=\{|0\rangle,|1\rangle,|2\rangle,|3\rangle\}$.
$\square$

\textbf{Remark 2.4.} Denote  $\langle \sigma_\imath \rangle_c=\mathrm{Tr}(\rho_c\sigma_\imath)$, $\langle \sigma_\imath \rangle=\mathrm{Tr}(\rho\sigma_\imath)$, $\imath=x,y,z$. From the relations $\langle\sigma_y\rangle=\mathrm{Tr}(\rho\sigma_y)=\mathrm{Tr}(U\rho_c U^\dag\sigma_y)=\langle \sigma_y \rangle_c$, $\langle \sigma_x \rangle=-\sin\alpha \langle \sigma_z\rangle_c+ \cos\alpha \langle\sigma_x \rangle_c$ and $\langle \sigma_z \rangle=\cos\alpha \langle \sigma_z\rangle_c+ \sin\alpha \langle\sigma_x \rangle_c~$, if the optimal approximation solution of $\rho=U\rho_c U^{\dag}$ with respect to the set $S^{'}$ can be expressed by $\langle \sigma_\imath\rangle $, then $\rho_c$ with respect to the set $S_1$ can also be expressed by $\langle \sigma_\imath \rangle_c$ duet to Lemma 2.3. Subsequently, we just need to discuss the optimal approximation of $\rho$ with respect to the set $S^{'}$.

\textbf{Lemma 2.5.} The following equality holds, $D_{S}(\rho)=\mathrm{min}\{\parallel\rho-\sum_i p_i\rho_i\parallel_1\}=\mathrm{min}\{2\sqrt{-\mathrm{Det}(\rho-\sum _{i}  p_i|\Psi_i\rangle\langle\Psi_i|)}\}$.

\emph{Proof.} Denote $A=\rho-\sum_{i} p_i\rho_i$. Obviously, $\mathrm{Tr }A=0$. Let $r_i$ be the eigenvalues of $A$. Then $ r_i=\pm\sqrt{|\mathrm{Det}A|}$, $i=1,2$.  Let $VAV^\dag=\mathrm{diag}(r_1,r_2)$. Then  we have $VAA^\dag V^\dag=\mathrm{diag}(r^2_1,r^2_2)$,
and  $\parallel\rho-\sum_{i} p_i\rho_i\parallel_1=2\sqrt{| \mathrm{Det}(\rho-\sum _{i}  p_i\rho_i )|}.$

Because $\mathrm{Det}(\rho-\sum _{i}  p_i\rho_i)=r_1r_2\leq0$,
 the original optimal convex approximation is reduced to finding the minimum value $\mathrm{min}\{2\sqrt{-\mathrm{Det}(\rho-\sum _{i}  p_i|\Psi_i\rangle\langle\Psi_i|)}\}$ such that $p_i\geq 0$, $\sum _{i}p_i=1$. $\square$

Concerning the optimal approximations to the model, we set
\begin{eqnarray}
&& F(p)=F(p_0,p_1,...,p_n)= \nonumber\\&&-\mathrm{Det}(\rho-\sum _{i}  p_i|\Psi_i\rangle\langle\Psi_i|)-\sum^n _{i=0}\lambda_i p_i-\lambda\sum^n _{i=0}p_i,
\end{eqnarray}
where $p_i\geq0,~\sum _{j}p_j=1$.
We first introduce the related KKT theorem \cite{Hoffmann}: Consider the optimization problem of finding $\mathrm{min} \{f(x)\}$ s.t. $g_i(x)\leq0$, $h_j(x)=0$, $i=1,2,...,m$, $j=1,2,...,k$. If point $x_0$ is regular (namely, $\nabla h_j(x_0)$, $j=1,...,k$, $\nabla g_i(x_0)$ linear independent), and $f(x_0)$ is a local minimum, then $\exists \lambda_i\geq0$, $\mu_j\in\mathbb{R}$ such that  $\nabla F(x_0)=0$, $\lambda_ig_i(x_0)=0,$ where $F(x)= f(x)+\Sigma^{m}_i\lambda_ig_i(x)+\Sigma^{k}_j\mu_jh_j(x)$.

The point $x$ satisfying the constraints and condition $\nabla F(x)=0,\lambda_ig_i(x)=0$ is called a KKT point. If $f(x)$ and $g_i(x)$ are both convex, and $h_j(x)$ are linear, then the point $x$ must be the solution of the optimization problem \cite{Forst16}.
Therefore, the optimization problem of $D_{S}(\rho)$ is equivalent to
\begin{eqnarray}
\nabla F=0,~\lambda_ip_i=0,~\lambda_i\geq0,~p_i\geq0,~\sum _{j}p_j=1
\end{eqnarray}
for $i=0,...,n$.

We first consider the optimal convex approximation of $\rho$ with respect to the set $S^{'}$. Denoting $u=\cos\theta,v=\sin\theta$, one obtains from (2)
\begin{eqnarray}
&&p_1+v^2p_2+u^2p_3+\lambda_0+\lambda-a=0, \nonumber
\\& &
p_0+u^2p_2+v^2p_3+\lambda_1+\lambda-1+a=0,\nonumber \\& &
v^2p_0+u^2p_1+p_3+\lambda_2+\lambda
-1+a+ \nonumber \\&&u^2+uv\langle \sigma_x \rangle-2u^2a=0,\nonumber \\& &
u^2p_0+v^2p_1+p_2+\lambda_3+\lambda-a-\nonumber \\&&
u^2-uv\langle\sigma_x\rangle+2u^2a=0,\nonumber \\& &
\lambda_ip_i=0,\lambda_i\geq0,p_i\geq0,i=0,1,2,3,\Sigma _ip_i=1.
 \end{eqnarray}
By solving the above equations, with similar detailed discussions and approaches to that in \cite{Theurer6}, we obtain the following complete analytical solutions.

\emph{\textbf{Tpye I-- Result of $\rho$ with respect to $S^{'}$:}}

\par \noindent  $i)$
If ${(1-\langle\sigma_z\rangle)}/{\langle\sigma_x\rangle} \geq\tan\theta\geq {\langle\sigma_x\rangle}/({1+\langle\sigma_z\rangle}),$
\begin{eqnarray}
D_{{S^{'}}}(\rho )= 2k \sqrt{a(1-a)}\sin\phi=\langle\sigma_y\rangle.
\end{eqnarray}
The corresponding coefficients associated with\\ $\{|0\rangle,|1\rangle,
|2\rangle,|3\rangle\}$ are given by
\begin{eqnarray}
&&p_0=\frac{1}{2}+\frac{\langle\sigma_z\rangle}{2}-\frac{\langle\sigma_x\rangle}{2}(\tan\theta)^{-1} -t,\nonumber
\\& &
p_1=\frac{1}{2}-\frac{\langle\sigma_z\rangle}{2}-\frac{\langle\sigma_x\rangle}{2}\tan\theta-t,\nonumber \\& &
p_2=\frac{\langle\sigma_x\rangle}{2}(\tan\theta+(\tan\theta)^{-1})+t,\nonumber \\& &
p_3=t,
\end{eqnarray}
where $t$ is a free parameter satisfying $\mathrm{min}\{\frac{{1}}{2}+\frac{{1}}{2}{\langle\sigma_z\rangle}-\frac{{1}}{2}{\langle\sigma_x\rangle}(\tan\theta)^{-1},
\frac{{1}}{2}-\frac{{1}}{2}{\langle\sigma_z\rangle}-\frac{{1}}{2}({\langle\sigma_x\rangle}\tan\theta)\}\geq t\geq0$. Denote $ A_1=\{\sum _{i}  p_i|i\rangle\langle i|\}$, then the minimum is attained for any $\rho^{opt}\in A_1 $.

\par \noindent  $ii)$
If $\tan\theta >{(1-\langle\sigma_z\rangle)}/{\langle\sigma_x\rangle} $, then
\begin{eqnarray}
D_{{S^{'}}}(\rho)=\sqrt{(\langle\sigma_x\rangle\sin\theta-(1-\langle\sigma_z\rangle)\cos\theta)^2+\langle\sigma_y\rangle^2},
\;
\end{eqnarray}
with optimal weights
\begin{eqnarray}
&&p_0=\frac{1}{2}+\frac{\langle\sigma_z\rangle}{2}-\frac{\langle\sigma_x\rangle}{2}(\tan\theta)^{-1}, \nonumber
\\& &
p_2=\frac{1}{2}-\frac{\langle\sigma_z\rangle}{2}+\frac{\langle\sigma_x\rangle}{2}(\tan\theta)^{-1}, \nonumber \\& &
p_1=p_3=0.
\;~~~~~~~~~~~~~~~~~~~~~~~~~~~~~~~~~~~~~~~~~~~~~
\end{eqnarray}
Set $ A_2=\{p_0\rho_0+ p_2\rho_2\}$. Then $S^{'}(\rho^{opt})=A_1\bigcup A_2$.

\noindent  $iii)$
If $\tan\theta < {\langle\sigma_x\rangle}/({1+\langle\sigma_z\rangle})$, then
\begin{eqnarray}
D_{{S^{'}}}(\rho)=\sqrt{(\langle\sigma_x\rangle\cos\theta-(1+\langle\sigma_z\rangle)\sin\theta)^2+\langle\sigma_y\rangle^2},
\;\nonumber
\end{eqnarray}
with optimal weights
\begin{eqnarray}
&& p_1=\frac{1}{2}+\frac{\langle\sigma_z\rangle}{2}+\frac{\langle\sigma_x\rangle}{2}\tan\theta, \nonumber
\\& &
p_2=\frac{1}{2}-\frac{\langle\sigma_z\rangle}{2}-\frac{\langle\sigma_x\rangle}{2}\tan\theta,\nonumber \\& &
p_0=p_3=0.
\;~~~~~~~~~~~~~~~~~
\end{eqnarray}
Denote $A_3=\{p_0\rho_0+ p_2\rho_2\}$. Then we have $S^{'}(\rho^{opt})=A_1\bigcup A_2\bigcup A_3$.

We now consider the optimal convex approximations of $\rho$ with respect to the set $S^{''}=S_2$, where $S_2=\{|\Psi_3\rangle,|\Psi_4\rangle,|\Psi_5\rangle,|\Psi_6\rangle\}$  with $|\Psi_3\rangle\equiv\cos\vartheta|0\rangle+\sin\vartheta|1\rangle$, $|\Psi_4\rangle\equiv\sin\vartheta|0\rangle-\cos\vartheta|1\rangle$, $|\Psi_5\rangle\equiv\sqrt{2}/2(|0\rangle+\sqrt{-1}|1\rangle)$  and $|\Psi_6\rangle\equiv\sqrt{2}/2(|0\rangle-\sqrt{-1}|1\rangle)$,  $\vartheta=\beta/2$, $\vartheta \in (0,\pi) $. Similar to the analysis for $\rho$ with respect to the set $S^{'}$, we obtain completely the analytical solutions of the optimal convex approximation of $\rho$ with respect to $S^{''}$.
For simplicity, denote $\mu=a+(\cos\vartheta)^2(1-2a)+k{\sqrt{a(1-a)}}\cos\phi \sin2\vartheta$ and $\nu=k{\sqrt{a(1-a)}}\sin\phi$.

\emph{\textbf{Tpye II-- Result of $\rho$ with respect to $S^{''}$}:}

$i)$ If $1-\nu\geq\mu\geq\nu $, then
\begin{eqnarray}
&& D_{{S^{''}}}(\rho )= \nonumber \\&& \sqrt{(\langle\sigma_x\rangle)^2+(\langle\sigma_z\rangle)^2-(\cos2\vartheta\langle\sigma_z\rangle+\sin2\vartheta\langle\sigma_x\rangle)^2}
,~~\end{eqnarray}
with the corresponding coefficients
\begin{eqnarray}\label{e9}
&&p_3=\mu-\nu-t, \nonumber
\\& &
p_4=1-\mu-\nu-t, \nonumber \\& &
p_5=2\nu+t,   \nonumber \\& &
p_6=t,
\end{eqnarray}
where $t$ is a free parameter such that $\min\{\mu-\nu,1-\nu-\mu-t\}\geq t\geq0$. Denote $ B_1=\{\sum^6 _{i=3}  p_i|\Psi_i\rangle\langle \Psi_i|\}$. Then the minimum is attained for any $\rho^{opt}\in B_1 $.

$ii)$ If $\mu>1-\nu$, then
\begin{eqnarray}
&& D_{{S^{''}}}(\rho )=\nonumber \\&& \sqrt{\langle\sigma_x\rangle^2+(\langle\sigma_y\rangle-1)^2+\langle\sigma_z\rangle^2
-(\langle U_{2\vartheta}\rangle+1-\langle\sigma_y\rangle)^2/2},~~~~~
\end{eqnarray}
where $\langle U_{2\vartheta}\rangle=\cos2\vartheta\langle\sigma_z\rangle+\sin2\vartheta\langle\sigma_x\rangle$,  and the optimal weights are given by
\begin{eqnarray}
&&p_3=\mu-\nu, \nonumber
\\& &
p_5=1-\mu+\nu,   \nonumber \\& &
p_4=p_6=0.
\;
\end{eqnarray}
Set $ B_2=\{p_3\rho_3+ p_5\rho_5\}$. Then $S^{'}(\rho^{opt})=B_1\bigcup B_2$.

$iii)$ If $\mu<\nu $, then
\begin{eqnarray}
&&D_{{S^{''}}}(\rho )= \nonumber \\&& \sqrt{\langle\sigma_x\rangle^2+(\langle\sigma_y\rangle-1)^2+\langle\sigma_z\rangle^2
-(\langle U_{2\vartheta}\rangle-1+\langle\sigma_y\rangle)^2/2}
 \nonumber
\end{eqnarray}
with the optimal weights
\begin{eqnarray}
&&
p_4=1-\mu-\nu,  \nonumber \\& &
p_5=\mu+\nu,   \nonumber \\& &
p_3=p_6=0.
\;
\end{eqnarray}
Let $B_3=\{p_4\rho_4+ p_5\rho_5\}$. We have $S^{'}(\rho^{opt})=B_1\bigcup B_2\bigcup B_3$.

Now let us analyze the above results. From the result of the case of $i)$ in \emph{Type (I)}, one observes that  the value of $D_{S ^{'}}(\rho)=\langle \sigma_y \rangle $ is independent of $\theta$.
That is to say, given $\rho$ that $\rho$ and $\theta$ satisfy the condition \emph{Type (I).i)},
the minimum distinguish ability (distance) from the target state to the
arbitrary approximate points does not very with $\theta$. $D_{S ^{'}}(\rho)$ only depends on the target state itself, which gives rise to a kind of invariance under the change of $\theta$. This distance invariance may play a role in judging the mask characteristics of mixed states \cite{Modi,Bin}. 
Secondly, the analytical results can be directly used to get the available states. For example, we can get all available states  which are only composed of convex combinations of Hadamard gate eigenstates by \emph{Type (I). i)}, as these states satisfy the condition $k=(1-2a)/(\sqrt {a (1-a)} (\cot\frac{\pi}{8}-\tan\frac{\pi}{8}))$ and $\phi=0$. 
The related probabilities are exactly $p_2=1-a+k\sqrt {a(1-a)}\tan\frac{\pi}{8}$ and $p_3=a-k\sqrt {a(1-a)}\tan\frac{\pi}{8}.$  In this respect, a more complete formula for the case of multiple bases can be obtained by the decomposition theorem in section III of this article. 
The analytical solution could be quite complicated and even difficult to analyze for the case of high-dimensional space or multi-variable parameters. However, similar to our optimal approximation model, one can get numerical solution for any required precision through the KKT theory and the optimization interval algorithm \cite{Ying Cui}.

At last, it easy to derive some interesting byproducts from the analytic solutions.
From (4) and (9), if both conditions of \emph{Type (I). i)} and \emph{Type (II). i)} hold, we have
\begin{eqnarray}
\langle\sigma_x\rangle^2+\langle\sigma_y\rangle^2+\langle\sigma_z\rangle^2
\equiv\langle U_{2\vartheta}\rangle^2+D_{S^{'}}(\rho)^2+D_{S^{''}}(\rho)^2.~~
\end{eqnarray}

From the above formulae, we can derive a triple uncertainty equality relation.
The triple uncertainty relation for three pairwise canonical observables, momentum $p$, coordinate $q$ and $r=-p-q$ satisfying $[p, q]=[q, r]=[r, p]=-i\hbar$, was derived by Kechrimparis and Weigert firstly in \cite{K}. By introducing the triple constant $\tau=\frac{2}{\sqrt{3}}$, the tight triple uncertainty relation $\Delta  p \Delta  q \Delta  r \geq (\frac{\tau}{2})^ {\frac{3}{2}}$ has been established. 
In stead of $p,q,r$, we consider the spin operators, $S_x=\frac{\sigma_x}{2}$, $S_y=\frac{\sigma_y}{2}$ and $S_z=\frac{\sigma_z}{2}$, which satisfy the commutator relations, $ [S_x, S_z]=-iS_y$, $[S_y, S_x]=-iS_z$ and $[S_z, S_y]=-iS_x$. $(S_x,S_y,S_z)$ consists a Schr\"{o}dinger triple consisting of three pairwise canonical observables.

Denote  $\Delta \Omega={\sqrt{\langle\Omega^2\rangle-\langle \Omega\rangle^2}}$. One has $\Delta S_x=\frac{{\sqrt{1-\langle \sigma_x \rangle^2}}}{2}$, $\Delta S_y=\frac{{\sqrt{1-\langle \sigma_y \rangle^2}}}{2}$ and $\Delta S_z=\frac{{\sqrt{1-\langle \sigma_z \rangle^2}}}{2}$.  From
$\langle\sigma_x\rangle^2+\langle\sigma_y\rangle^2+\langle\sigma_z\rangle^2
=4k^2a(1-a)+(1-2a)^2\equiv f_1(k,a),$
we get \begin{eqnarray}(\Delta S_x)^2+(\Delta S_y)^2+(\Delta S_z)^2=\frac{3-f_1(k,a)}{4}.\end{eqnarray}
On the other hand,
 \begin{eqnarray}&&|\langle S_x\rangle|+|\langle S_y\rangle|+|\langle S_z\rangle|
\leq \nonumber \\&&\frac{1}{2}-a+k\sqrt{a(1-a)}(\sin\phi+\cos\phi)\leq  \nonumber \\&& \frac{1}{2}-a+k\sqrt{2a(1-a)}\equiv f_2(k,a).\end{eqnarray}

Set 
\begin{eqnarray}\lambda=\mathrm{max}\{\frac{4f_2(k,a)}{3-f_1(k,a)}\},
\end{eqnarray} 
where $k\in[0,1]$, $a\in[0,\frac{1}{2}]$. 
A simple calculation gives $\lambda={\sqrt{3}}$ when $k=1$ and $a=\frac{1}{2}-\frac{\sqrt{3}}{6}$. Therefore, we have the following triple uncertainty relation:
\begin{eqnarray}
(\Delta S_x)^2+(\Delta S_y)^2+(\Delta S_z)^2\geq\frac{\tau}{2}(|\langle{S_x}\rangle|
+|\langle{S_y}\rangle|+|\langle{S_z}\rangle|),
\end{eqnarray}
where $\tau=\frac{2}{\sqrt{3}}$ is a triple constant appeared also in \cite{K}. 
The equality in (18) holds when $a=\frac{1}{2}-\frac{\sqrt{3}}{6}$, $k=1$ and $\phi=\frac{\pi}{4}$. The bound $\frac{\tau}{2}$ is tight, which is in consistent with the result in \cite{K,Gisin17}.

Denote $M_0=\frac{{\sqrt{1-\langle U_2\vartheta \rangle^2}}}{2}$, $M_1=\frac{{\sqrt{1- D_{S^{'}}(\rho)^2}}}{2}$ and $M_2=\frac{{\sqrt{1- D_{S^{''}}(\rho)^2}}}{2}$. Let  $\rho=diag( \frac{1}{2}-\frac{\sqrt{3}}{6},1,\frac{\pi}{4})$. From (14) we have a triple uncertainty equality relation:
\begin{eqnarray}
( M_0)^2+(M_1)^2+(M_2)^2\equiv\frac{\tau}{2}(|\langle{S_x}\rangle|
+|\langle{S_y}\rangle|+|\langle{S_z}\rangle|).
\end{eqnarray}
By the conditions of \emph{Type(I).i)} and \emph{Type(II).i)},  for all $\theta\in (0.3509,0.6319)$ and  $\vartheta\in  (0.9061,1.4501)\cup (2.4768, 3.0209)$, the formula (19) holds.
(19) can be experimentally verified by measuring $M_1$ and $M_2$, where the
parameter $\theta$ or $\vartheta$ is easy to adjust physically, since 
they could be the obliquely angles of the polarizer or the half-wave plate with respect to the horizontal axis in optical experiments. For instance, if the optical axis of the half-wave plate is obliquely placed to $\vartheta$ angle with the horizontal axis, it constitutes the $U_{2\vartheta}$ logic gate \cite{Barnett13}. It is of great significance to manipulate such physical quantities in quantum physics  \cite{Robert,Hengyan}.

\section{  decompositions of mixed states to three quantum logic gates }
We now study the decomposition of a given qubit state with respect to the set of eigenstates of   three real different quantum logic gates. Without loss of generality, we choose $S=\{|\Psi_1\rangle,|\Psi_2\rangle,|\Psi_3\rangle,
|\Psi_4\rangle,|\Psi_5\rangle,|\Psi_6\rangle\}$. We consider the optimal convex approximations of the initial states $\rho_c$ with respect to the set $S$.

We further denote $|2\rangle=(\cos\theta|0\rangle+\sin\theta|1\rangle)$, $|3\rangle=(\sin\theta|0\rangle-\cos\theta|1\rangle)$, where $\theta=\frac{(\beta-\alpha)}{2}, \beta>\alpha$. Let
$|4\rangle=\frac{\sqrt{2}}{2}(|0\rangle+\sqrt{-1}|1\rangle)=|\Psi_5\rangle$ and $|5\rangle=\frac{\sqrt{2}}{2}(|0\rangle-\sqrt{-1}|1\rangle)=|\Psi_6\rangle$.
Denote $S^{'''}=\{|0\rangle,|1\rangle,|2\rangle,|3\rangle,
|4\rangle\,|5\rangle\}$, where $|2\rangle ,..., |5\rangle$ are two-dimensional superposition states.

\textbf{Lemma 3.1.} \emph{The problem of the optimal convex approximations of $\rho_c$ with respect to  $S$ is equivalent to that of $\rho$ with respect to $S^{'''}$.}

\emph{Proof.}
Suppose $U$ is the same isometric transformation as in Lemma 2.3.. One has 
$$
U|\Psi_1\rangle=|0\rangle,~ U|\Psi_2\rangle=-|1\rangle,~
U|\Psi_3\rangle\equiv|2\rangle,~ U|\Psi_4\rangle\equiv|3\rangle.
$$
It is direct to verify that
\begin{eqnarray}
&&U|\Psi_5\rangle\langle\Psi_5|U^\dag=\nonumber \\&&\left(
\begin{array}{cc} \cos\frac{\alpha}{2}
& \sin\frac{\alpha}{2}\\
-\sin\frac{\alpha}{2}
& \cos\frac{\alpha}{2}\\ \end{array} \right )
\left(
\begin{array}{cc} \frac{1}{2}
& -\frac{i}{2}\\
\frac{i}{2}
& \frac{1}{2}\\ \end{array} \right )
\left(
\begin{array}{cc} \cos\frac{\alpha}{2}
& -\sin\frac{\alpha}{2}\\
\sin\frac{\alpha}{2}
& \cos\frac{\alpha}{2} \\ \end{array} \right )\nonumber \\&& =|4\rangle\langle4|
\end{eqnarray}
and  $U|\Psi_6\rangle\langle\Psi_6|U^\dag=U|5\rangle\langle5|U^\dag=|5\rangle\langle5|.$
Therefore, we obtain
 \begin{eqnarray}& &\parallel U(\rho_c-\sum^{6}_{i=1} p_i\rho_i )U^\dag\parallel=\parallel U\rho_c U^\dag-\sum^{6}_{i=1} p_i U|\Psi_i\rangle\langle\Psi_i|U^\dag\parallel \nonumber \\& &
 =\parallel \rho-\sum^{5}_{i=0} p_i |i\rangle\langle i|\parallel,\end{eqnarray}
which proves that the optimal convex approximations of $\rho_c$ with respect to the set $S$ is equivalent to that of $\rho$ with respect to the set $S^{'''}$. $\square$

\textbf{Theorem 3.2.} \emph{ A mixed state $\rho$ can be decomposed via the vectors in  $S^{'''}$ or $S$ if and only if $(1-\langle \sigma_y \rangle)^2\geq\langle \sigma_x \rangle^2+\langle \sigma_z \rangle^2$.}

\emph{Proof.}~~ That a mixed state can be decomposed by $S^{'''}$ means that $D_{S^{'''}}(\rho )=0$. From our optimal approximation model it is easy to get $D_{S^{'''}}(\rho )=0 \Longleftrightarrow$
\begin{eqnarray}
&& p_1+(1-\cos^2\theta)p_2+ \cos^2\theta p_3+\frac{1}{2}p_4+\frac{1}{2}p_5\nonumber\\&&-\frac{1}{2}+\frac{1}{2}\langle \sigma_z \rangle=0, \nonumber
\\& &
(1-\cos^2\theta)p_0+\cos^2\theta p_1+p_3+\frac{1}{2}p_4+\frac{1}{2}p_5\nonumber\\&&-\frac{1}{2}-\frac{1}{2}\langle \sigma_z \rangle+\langle \sigma_z \rangle\cos^2\theta+\frac{1}{2}\langle \sigma_x \rangle\sin2\theta=0, \nonumber \\& &
\frac{1}{2}p_0+\frac{1}{2}p_1+\frac{1}{2}p_2+\frac{1}{2}p_3+p_5\nonumber\\&&-\frac{1}{2}+\frac{1}{2}\langle \sigma_y \rangle=0,\nonumber \\& &
\Sigma^5 _{i=0}p_i=1,~p_i\geq0,~i=0,1,...,5.
\end{eqnarray}
By solving the above equations, we obtain
\begin{eqnarray}
&& p_0=\frac{1}{2}+\frac{1}{2}\langle \sigma_z \rangle-\frac{1}{2}\langle \sigma_x \rangle\tan^{-1}\theta-\frac{1}{2}\langle \sigma_y \rangle-c_1-c_2,\nonumber\\&&
p_1=\frac{1}{2}-\frac{1}{2}\langle \sigma_z \rangle-\frac{1}{2}\langle \sigma_x \rangle\tan\theta-\frac{1}{2}\langle \sigma_y \rangle-c_1-c_2,\nonumber\\&&
p_2=\frac{1}{2}\langle \sigma_x \rangle \tan^{-1}\theta+\frac{1}{2}\langle \sigma_x \rangle\tan\theta+c_2,\nonumber\\&&
p_3=c_2,\nonumber\\&&
p_4=\langle \sigma_y \rangle+c_1,\nonumber\\&&
p_5=c_1,~\mathrm{and}~ p_i\geq0,~ i=0,1,...,5,
\end{eqnarray}
From that $c_1$ and $c_2$ are arbitrary non-negative arguments such that $p_i\geq0$, we have
\begin{eqnarray}
&&1+\langle \sigma_z \rangle-\langle \sigma_x \rangle\tan^{-1}\theta-\langle \sigma_y \rangle\geq0, \nonumber\\&&1-\langle \sigma_z \rangle-\langle \sigma_x \rangle\tan\theta-\langle \sigma_y \rangle\geq0.
\end{eqnarray}
Since $0\leq \langle \sigma_{x}\rangle, \langle \sigma_{y}\rangle,\langle\sigma_{z}\rangle\leq 1$, we get
\begin{eqnarray}
\frac{1-\langle \sigma_z \rangle-\langle \sigma_y \rangle}{\langle \sigma_x \rangle}\geq\tan\theta\geq
\frac{\langle \sigma_x \rangle}{1+\langle \sigma_z \rangle-\langle \sigma_y \rangle}.
\end{eqnarray}
Therefore, $(1-\langle \sigma_y \rangle)^2\geq\langle \sigma_x \rangle^2+\langle \sigma_z \rangle^2$. Conversely,
there must also be $\theta$ such that (23) holds, and the mixed state can be decomposed by  $S^{'''}$.

Noticing $\langle \sigma_x \rangle=-\sin\alpha \langle \sigma_z\rangle_c+ \cos\alpha \langle\sigma_x \rangle_c$,
$\langle \sigma_z \rangle=\cos\alpha \langle \sigma_z\rangle_c+ \sin\alpha \langle\sigma_x \rangle_c$ and $\langle\sigma_y \rangle=\langle\sigma_y \rangle_c,$
we can easily deduce $(1-\langle \sigma_y \rangle_c)^2\geq\langle \sigma_x \rangle_c^2+\langle \sigma_z \rangle_c^2.$
Combining with formula (21), we have that $\rho_c$ can be also decomposed by $S$ if and only if $(1-\langle \sigma_y \rangle_c)^2\geq\langle \sigma_x \rangle_c^2+\langle \sigma_z \rangle_c^2$.
$\square$

In \cite{Sacchi41,Theurer6}, three special quantum logic (X,Y,Z)-gates are considered.
A mixed state can be decomposed if and only if the condition $a-k\sqrt{a(1-a)}\cos \phi-k\sqrt{a(1-a)}\sin\phi\geq0$ holds, namely, 
$$
1-\langle \sigma_y \rangle\geq\langle \sigma_x \rangle+\langle \sigma_z \rangle~~~(\star).
$$
Obviously, the above condition $(\star)$ implies $(1-\langle \sigma_y \rangle)^2\geq\langle \sigma_x \rangle^2+\langle \sigma_z \rangle^2~~~(\star\star)$. Conversely, if condition $(\star\star)$ holds, from (25) $(\star)$ holds by taking $\theta =\frac{\pi}{4}$.
In fact, when $(\star\star)$ holds, $\theta$ may take different values, namely, the decompositions of mixed states are not unique. In addition, if the decomposability condition holds, we can get all possible decompositions from formula (23) by choosing a suitable angle. By Theorem 3.2. it is also easy to get all   optimal approximations of mixed state $\rho$ to $S^{'''}$.
The three Pauli matrices case considered in \cite{Sacchi41,Theurer6} is just a special case of $S^{'''}$ when $\theta = \frac{\pi}{4}$ in section III of this paper.
That is to say, our results fully contain the corresponding conclusions in \cite{Sacchi41,Theurer6}.

\section{Conclusion}

In summary, we have obtained the explicit formulae of optimal approximations for arbitrary parameter $\theta$ or $\vartheta$  about a target qubit state to the eigenstate set of real quantum logic gates.  From the analytic formulae of minimum distance expressed in terms of the average values of Pauli operators, we also obtained an interesting trade-off relation about uncertainty. In addition, we have shown a simple inequality criterion, from which it is easy to judge whether a given state can be decomposed by three quantum logic gates. Our results give a way to search for desired available convex combinations by changing $\theta$ or $\vartheta$, which may be used as significant tools in quantum cryptography, clone,  and coherence field \cite{Bae,Scarani18,Bergou19,Baumgratz20}. More importantly, the model and solution, the idea of isometric transformation and  mathematical methods provided in this paper may be applied to a wider field of quantum information and computing.

\bigskip
\noindent {\bf Acknowledgments} 
This work is supported by NSFC (11765016,11847108,11675113), Beijing Municipal Commission of Education (KZ201810028042), Scientific research project of Jiangxi Provincial Department of Education  under No. (GJJ12607, GJJ190888), Beijing Natural Science Foundation (Z190005),
and Academy for Multidisciplinary Studies, Capital Normal University.

\end{document}